\documentstyle[aps,fleqn]{revtex}

\newenvironment{figures}[1]%
{\begin{list}{}{\settowidth{\labelwidth}{#1}
  \setlength{\leftmargin}{\labelwidth}
  \addtolength{\leftmargin}{\labelsep}
  \setlength{\parsep}{1ex plus0.7ex minus0.7ex}
  \setlength{\itemsep}{0.8ex}
  }}{\end{list}}

\begin{document}

\title {{\small
\hspace{11cm} NIKHEF 95--033  \\ 
\hspace{11cm} hep-ph/9507394 }\\ \vspace{1cm}
\bf Production of vector mesons by real and virtual
photons at high energies}

\author{
L.P.A. Haakman
$^{a}$,
A. Kaidalov 
$^{b}$,
J.H. Koch
$^{a}$}

\address{ 
$^{a}$
National Institute for Nuclear Physics 
and High Energy Physics (NIKHEF--K), \\
P.O. Box 41882, NL-1009 DB Amsterdam, The Netherlands \\ 
$^{b}$ 
Institute of Theoretical and Experimental Physics, \\
B. Cheremushinskaya 25, 117 259 Moscow, Russia}

\date{24 July 1995}

\maketitle

\begin{abstract}
A model based on Reggeon field theory is used for the description
of the photo- and electroproduction of vector mesons 
($\rho^0, \omega, \phi, J /\psi$)
on the proton. Its main feature is the dependence of the Pomeron 
intercept on the virtuality of the photon. 
The few parameters of the model were 
determined by fitting the total cross sections for real and virtual 
photons on the proton. This allows a parameter-free description 
of the energy
dependence of vector meson production. Very good agreement with the
existing high energy data is obtained.
\end{abstract}


The study of the production of vector mesons 
($\rho^0 ,\omega, \phi, J /\psi$) by real and virtual photons at 
high energies provides important information on the diffractive 
mechanism at very high energies. In particular, it allows 
to study the transition from the nonperturbative regime 
at low values of $Q^2$, the negative of the 
four momentum of the photon squared, to the perturbative 
regime at large $Q^2$.
Many recent papers \cite {1,2,3,4,5,6} have addressed the problem of
the production of vector 
mesons. It was mainly described by the exchange of 
a `soft' Pomeron \cite {1,2}. At large $Q^{2}$ this 
reaction was seen as a  test of perturbative QCD and 
of the exchange of the so-called `BFKL Pomeron' \cite {3,4,5,6}.
The new HERA data \cite {7,8} showed a non-trivial energy
 dependence of the production of $\rho^0$ and $\psi$ mesons. 
The photoproduction cross section for $\rho^0$ mesons increases 
slowly with energy, consistent with the soft Pomeron
model of Ref. \cite {1} . However, at high $Q^{2}$ the increase 
is much faster than predicted by this model. 
This faster increase with energy is also seen
for the production of $J /\psi$--mesons. How these high $Q^{2}$
data agree with the perturbative approach 
or the BFKL model is not certain yet.

In this note we extend the approach in Ref.\cite {9} to the 
production of vector mesons. This model was proposed 
for the description of the total 
cross section for the absorption of real and virtual 
photons at high energies. The central idea is that the
Pomeron singularity, which governs the 
high energy behavior of the diffractive process,
is not a simple pole characterized
by a fixed intercept at $t=0$, $ \alpha_{P} (0)$. 
Instead, multiple Pomeron exchanges must also be considered, 
leading to modifications of the simple pole picture. These multiple 
exchanges are especially important for the `supercritical' 
Pomeron, {\it i.e.} when $\Delta = \alpha_{P} (0) - 1 > 0$.
 Single as well as multiple Pomeron 
exchange are actually needed to obtain an amplitude
satisfying the requirements of unitarity. Diagrammatic 
techniques for Reggeon exchanges \cite {10} and the AGK cutting
rules \cite {11} allow one to calculate the contributions 
of  many such Pomeron exchanges to the scattering amplitudes 
and relate them to the properties of multiparticle production.
Extensive phenomenological studies based on
this principle (see {\it e.g.} Refs.[12] and [13]) 
have shown that all data on soft 
hadronic interactions can succesfully be described. 
It  follows from these analyses that the observed intercept,
 $\alpha^{eff}_P(0) = 1.08$, which characterizes the energy 
dependence of hadronic cross sections in the currently  available 
range of energies,
$\sigma_{hp}^{tot} \sim s^{\alpha^{eff}_{P} (0) - 1}$,
is substantially smaller than the intercept, $\alpha_P(0)$, of 
the single Pomeron pole itself. In the studies \cite {12,13} 
which take into account sequential (`eikonal') many 
Pomeron exchanges, one finds $\alpha_P(0) =1.12 - 1.15$ .
Taking into account 
a more complete set of diagrams, which also include interactions 
among the exchanged Pomerons, an even larger intercept
of $\alpha_P(0) \approx 1.2$ for the single pole 
is obtained \cite {14}. So the {\it effective} intercept, 
 which determines the overall energy dependence
of the scattering amplitude, is due to the contribution 
from the single  Pomeron exchange as well as of the Pomeron cuts, 
the exchanges of many Pomerons. It was argued in 
Ref. \cite {9} that the relative importance of these 
cuts strongly decreases with the virtuality. 
In an electromagnetic reaction, real photons probe the entire
complex of single and multiple Pomeron exchanges. However, with 
higher virtuality the space-time resolution increases and 
one `sees' the single Pomeron exchange. This means that the 
intercept depends on $Q^2$, {\it i.e.} $\alpha^{eff}_P (0,Q^2)$. 
The exponent that governs the energy dependence of 
the total photon cross section thus also changes with $Q^2$, 
$\alpha_{P}^{eff} (0,Q^2) -1 \equiv \Delta^{eff}(Q^2)$.
For real photons, at $Q^2=0$, it has the value
$\Delta^{eff}(0)\approx 0.08$ which also
governs the soft hadronic interactions.
At high $Q^2$, $Q^2 \gg 1$ GeV$^2$, 
 we expect $\Delta^{eff}\approx 0.24$ due to the Pomeron pole.
The parametrization chosen in Ref.\cite{9} was therefore
\begin{equation}
\Delta^{eff}(Q^{2}) = 
\Delta(0)~\left(1 + \frac{2~Q^{2}}{d + Q^{2}}\right)
\label{delta}
\end{equation}
with $\Delta (0) = 0.077$ and $d = 1.117$ GeV$^{2}$.
At  larger values of $Q^2$
it is necessary to take into account effects due to QCD 
evolution,  which lead to an additional increase of 
the energy dependence with $Q^{2}$. 

The approach sketched above was applied to the 
electromagnetic structure function of 
the nucleon \cite{9} and we compare it here to the 
new H1 data \cite{18}.
For values of $Q^{2}$ above $2$ GeV$^{2}$
perturbative QCD evolution was applied  on the two loop level.
For the virtual photon-proton total cross section the convention
is adopted that
\begin{equation}
\sigma_{\gamma^{*}p}=
\frac{4 \pi^2 \alpha_{\scriptstyle{EM}}}
{Q^2 (1-x)} F_{2}(x,Q^2) \label{defcrs} ~~
\end{equation}
where $x$ is related to the total centre-of-mass (CM) 
energy $W$ by
\begin{equation}
W^2 \equiv s = \frac{Q^2}{x}(1-x)+m_{p}^2 ~~
\end{equation}
with $m_p$ the mass of the proton.
The results of the model from Ref.\cite{9} are shown in 
Fig.1, together with the predictions for the new H1 data.
The agreement with the combined experimental data is excellent.
The steady increase of $\Delta^{eff}$ with $Q^{2}$
is clearly seen. As the $Q^{2}$ dependence of $\Delta^{eff}$
dies off with increasing virtuality in this model, the change 
in slope for values of $Q^{2} \gg d$   is due to the QCD 
evolution. We also show a cross section  for a low 
virtuality, $Q^2=0.35$ GeV$^2$, 
which can be compared with the forthcoming data from E665. 
The preliminary data \cite{19} indicate that our model also
works very well in the kinematical range covered by E665.  

The same approach can be used for other diffractive processes,
such as the production of vector mesons by 
real as well as virtual photons, $\gamma^{(*)} p \rightarrow V p$.
The combined contribution of the Pomeron pole and the cuts to the 
amplitude for this process can be written in the following form
\begin{equation}
T(s,t,Q^{2},m_{V}^{2}) = 
f(Q^{2},m_{V}^{2},t) \left(\frac{s}{s_{0}}\right)^
{\alpha_{P}^{eff}(t, \bar{Q}^{2})} \eta(\alpha_{P}^{eff})
\label{amplitude}
\end{equation}
 where $\eta(\alpha)$ is the signature factor and the 
dependence on the total CM--energy 
is determined by the effective Pomeron trajectory with
$\alpha_{P}^{eff}(t, \bar{Q}^{2})$.
 We have indicated through the argument $\bar{Q}^2$
that this effective parameter depends on the characteristic 
virtuality of the process. 
For the production of states made up of light quarks ($u,d,s$) 
we have $\bar{Q}^{2} = Q^{2}$, 
while for vector mesons made up of heavy quarks we choose 
\begin{equation}
\bar{Q}^{2} = c~m_{q}^{2} + Q^{2}
\end{equation}
where $m_{q}$ is the mass of the heavy  quark and $c \sim 1$.
In this way we take into account that for production of states
made up of heavy quarks the  contribution of cuts is small 
already at low $Q^{2}$. The effective pomeron 
trajectory $\alpha^{eff}_{P} (t,\bar{Q }^{2})$ is   
assumed to be linear in $t$, {\em i.e.}
\begin{equation}
\alpha^{eff}_{P} (t,\bar{Q}^{2}) = 1 + \Delta^{eff}(\bar{Q}^{2}) + 
\alpha'_{P} t
\label{intercept}
\end{equation}
with the same parametrization for $\Delta^{eff}(\bar{Q}^2)$, 
Eq.(\ref{delta}), as in the structure function 
of the proton in Ref. \cite {9}. 
We take $\alpha'_{P} = 0.25$ GeV$^{- 2}$,
which was obtained from the analysis of hadronic interactions. 
In principle $\alpha'_{P}$ also 
depends on $\bar{Q}^{2}$, but from analysis of elastic hadronic
reactions it follows that this dependence is rather weak
and can be neglected when discussing the energy dependence of the 
cross section. The t--dependence of the function 
$f(Q^{2},m_{V}^{2},t)$ is well known from {\it e.g.} 
proton scattering, $\pi N$ scattering and the production 
of $\rho^0$ mesons on the proton at low energies. It is taken to 
be exponential \cite{1},
\begin{equation}
f(Q^{2},m_{V}^{2},t) = \tilde{f}(Q^{2}, m_{V}^{2}) \exp(R^{2}t) 
\end{equation}
with
$R^{2} = R^{2}_{p} +  R^{2}_{V}(Q^{2})$ where  
\begin{eqnarray}
R^{2}_{V} &=& 
\frac{R^{2}_{0V} m_{\rho}^{2}}{m_{V}^{2} + Q^{2}} \nonumber \\
R^{2}_{p} &=& 2~{\rm GeV}^{-2} \nonumber \\
R^{2}_{0V} &=& R^{2}_{\rho} = 1~{\rm GeV}^{-2} ~~.
\end{eqnarray}
It has a factorizable form and takes into account the change 
with $Q^{2}$ and
$m_{V}^{2}$ of the radius of the $\gamma V  P$ vertex.

The contribution of the Pomeron to the differential cross 
section for the  $\gamma^{(*)} p \rightarrow V p$ 
reaction then becomes
\begin{equation}
\frac{d\sigma}{dt} = 
\frac{1}{16 \pi s^{2}} | T(s,t,Q^{2}, m_{V}^{2})|^{2} = 
 {\cal F}(Q^{2},m_{V}^{2}) 
\left(\frac{s}{s_{0}}\right)^{2 \Delta^{eff} 
(Q^{2})} \exp (\Lambda (s) t)
\label{diffVMcrs}
\end{equation}
where 
\begin{equation}
{\cal F}(Q^{2}, m_{V}^{2}) =\frac{|\tilde{f}(q^{2},m_{V}^2) 
\eta(\alpha_{P})|^{2}}{16 \pi}
\end{equation}
and
\begin{equation}
\Lambda(s) = 2\left(R^{2} + \alpha'_{P}
\ln (\frac{s}{s_{0}})\right)
\end{equation}
is the slope of the diffraction cone. For the scale 
factor, $s_{0}$, we choose $s_{0} = m_{V}^2 + Q^{2}$. 
This can be done, since the function ${\cal F}(Q^{2}, 
m_{V}^{2})$ is unknown and will be fitted to the data. 
The dependence on the energy $s$  is however entirely determined
from Eqs.(\ref{delta}) and (\ref{diffVMcrs}). 
The cross section due to Pomeron exchange, 
Eq.~(\ref{diffVMcrs}), can then be written as
\begin{equation}
\frac{d\sigma}{dt} = 
\Phi(Q^{2}, m_{V}^{2}) (F_{2,p}^{sea} (x,\bar{Q}^2)^{2}
\exp(\Lambda (s) t)
\label{diffcrs}
\end{equation}
where $F_{2,p}^{sea} (x,\bar{Q}^2)$ is the sea quark 
contribution to the
structure function of the proton \cite {9} with $x = \bar{Q}^{2}/({\bar{Q}^2+s})$.
Eq.(\ref{diffcrs}) allows us to include
perturbative QCD corrections, which are important at 
large values of $Q^2$, by applying the QCD evolution
equations to the structure function.
For the production of vector mesons made up 
of $u$ and $d$ quarks ($\rho^0 , \omega$), 
at not too large energies also secondary 
exchanges --- the ${\it f}$ and $A_{2}$ Regge poles 
--- contribute in addition to the Pomeron exchange. 
In the parametrization of the structure function
$F_{2}^{p} (x,Q^2)$  in Ref. \cite {9} they correspond to 
the valence quark contribution. In the following we shall
assume that for $(\rho^0,\omega)$--production the ratio
of the Pomeron to secondary exchanges is the same as for the 
structure function. This means that the entire structure
function $F_{2}^p$ as parametrized in Ref.\cite{9} enters into Eq.(\ref{diffcrs}) 
for vector meson production. 

Results for the total cross section for the  reaction 
$\gamma^* p \rightarrow \rho^0 p$, obtained by 
integrating Eq.(\ref{diffcrs}) over t,
are shown in Fig.2. At each value of $Q^{2}$, 
the curves are normalized to obtain a good fit to the data.
For energies where we expect our Reggeon-based model to apply, 
$s > 10$ GeV$^2$, the data are described well. 
Especially the strong increase of the cross section with energy
that was observed by the ZEUS Collaboration 
\cite {7} at $Q^{2} \sim 10$ GeV$^{2}$ is well reproduced 
by the model. It will be interesting to test the 
model predictions for high energies,
in particular at lower values of $Q^{2}$.

The data for $\omega$--meson photoproduction are shown 
in Fig.3 . Only relatively low energy data are available 
and their $W$ dependence is well described.
For the photoproduction of $\phi$ and $ J/\ \psi$ mesons 
there is no valence quark contribution 
and only Pomeron exchange contributes to the 
amplitude. Using for the charmed quark mass 
$m_{c}^{2} = 2.6$ GeV$^{2}$,
we obtain good agreement with recent HERA results 
\cite {8} as shown in Fig.3 .
The energy dependence of the $\phi$ meson 
is weaker and is also well 
reproduced by the model.

In conclusion, the data on photo- and electroproduction 
of vector mesons yield additional support for the idea 
of a steady change of the effective intercept of the 
Pomeron singularity. The simple model based on this idea
can be used to predict the energy dependence for other
diffractive processes in photo- and electroproduction at 
high energies.

\vspace*{1cm}
{\bf  ACKNOWLEDGEMENT}

The work of L.H. and J.K. is part of the research program 
of the Foundation for Fundamental
Research of Matter (FOM) and the National Organization for 
Scientific Research (NWO). The collaboration with ITEP
is supported in part by a grant from NWO
and by grant 93-79 of INTAS. A.K. also acknowledges support
from grant J74100 of the International Science Foundation 
and the Russian Goverment.

\small{

}

\newpage

\section*{\bf Figure Captions}

\begin{figures}{3}

\item[{\bf Fig.1}] The total cross sections 
for real and virtual photons
for the proton  obtained from the model of Ref.\cite{9}. 
Reading from top to bottom the curves  belong to :\ $Q^2= 0,~~0.35,~~1.4,~~3,~~8.5,~~15,~~35,~~65,~~125$ GeV$^2$. 
The H1 data \cite{18} are denoted by the black squares
, the ZEUS data \cite{15} by the open circles
and the low energy data from BCDMS, SLAC and NMC \cite{15} 
by the open triangles. The real photon  data [16--17] are
indicated by the open squares.

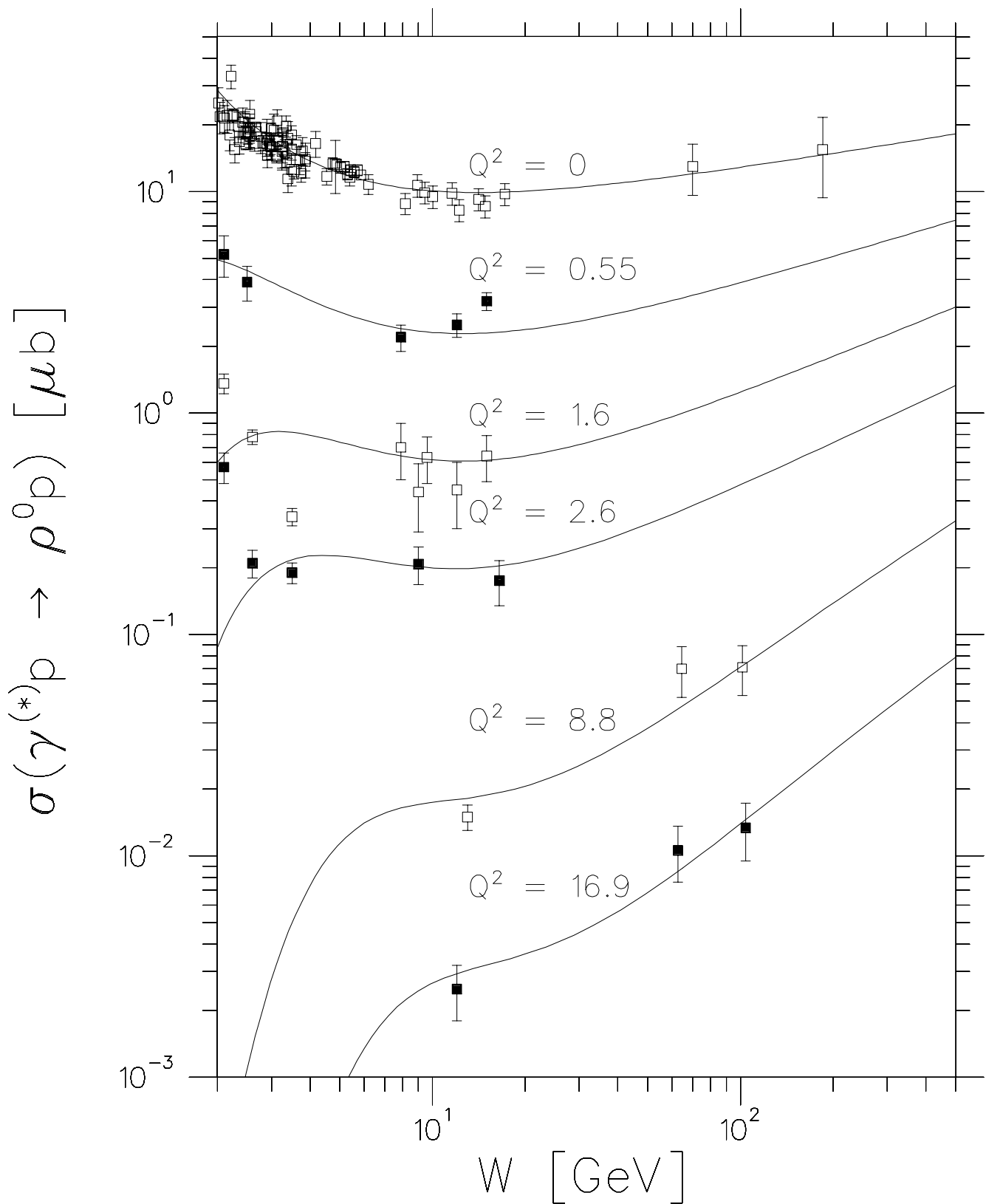
\item[{\bf Fig.2}] The $\gamma^{*}p\rightarrow \rho^0 p$ 
cross sections for several values of $Q^2$ obtained from 
our model. The data for $W > 50$ GeV at high $Q^2$ are
 the recent ZEUS--data.
All data are from \cite{7} and references therein.

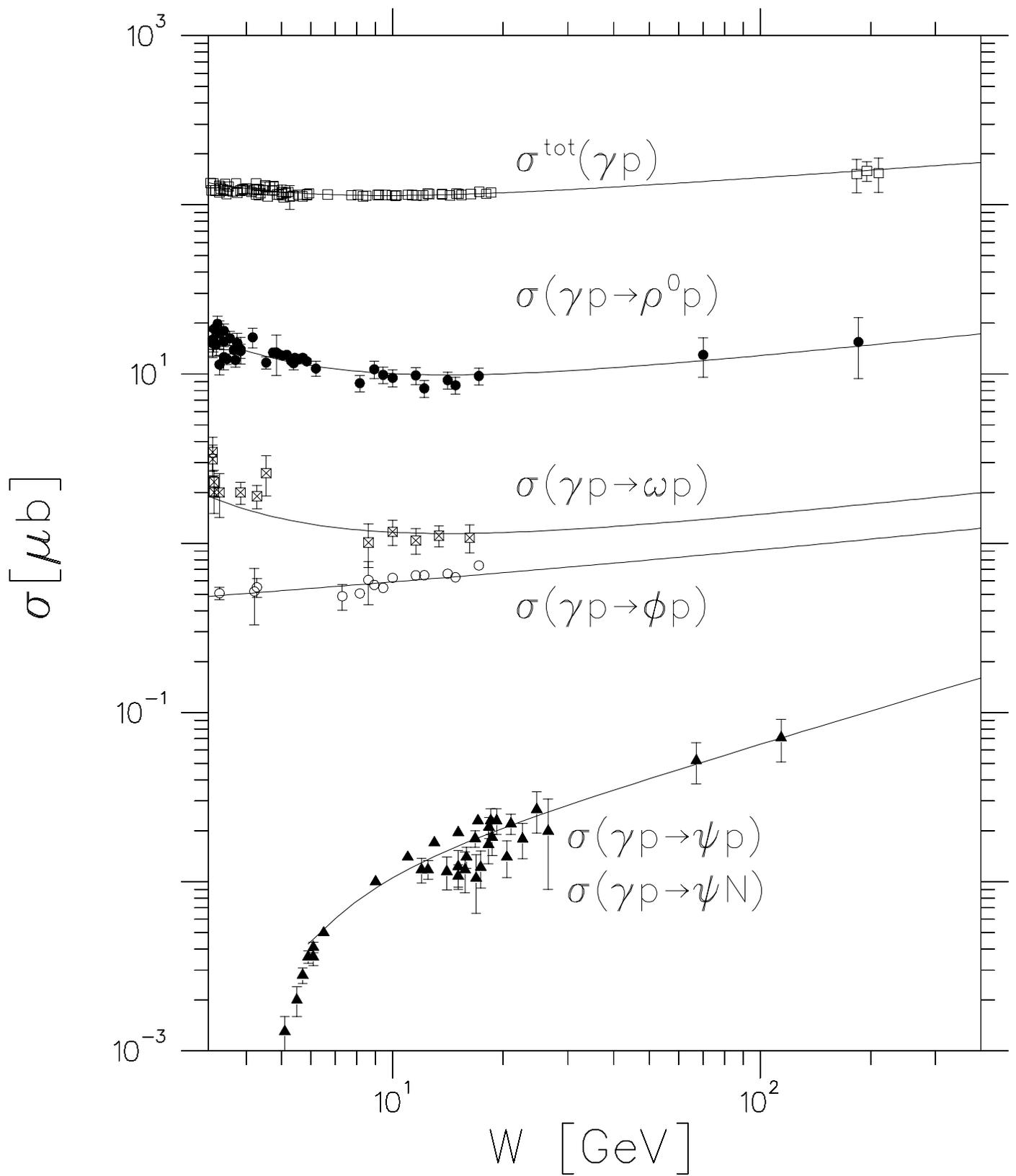
\item[{\bf Fig.3}] A compilation of total photoproduction 
and vector meson production cross sections  resulting from 
the model.
The  high energy data are from Refs.\cite{7}, \cite{8} 
and \cite{17}. The low energy data are from \cite{16} and 
references in \cite{8}.

\end{figures}
 
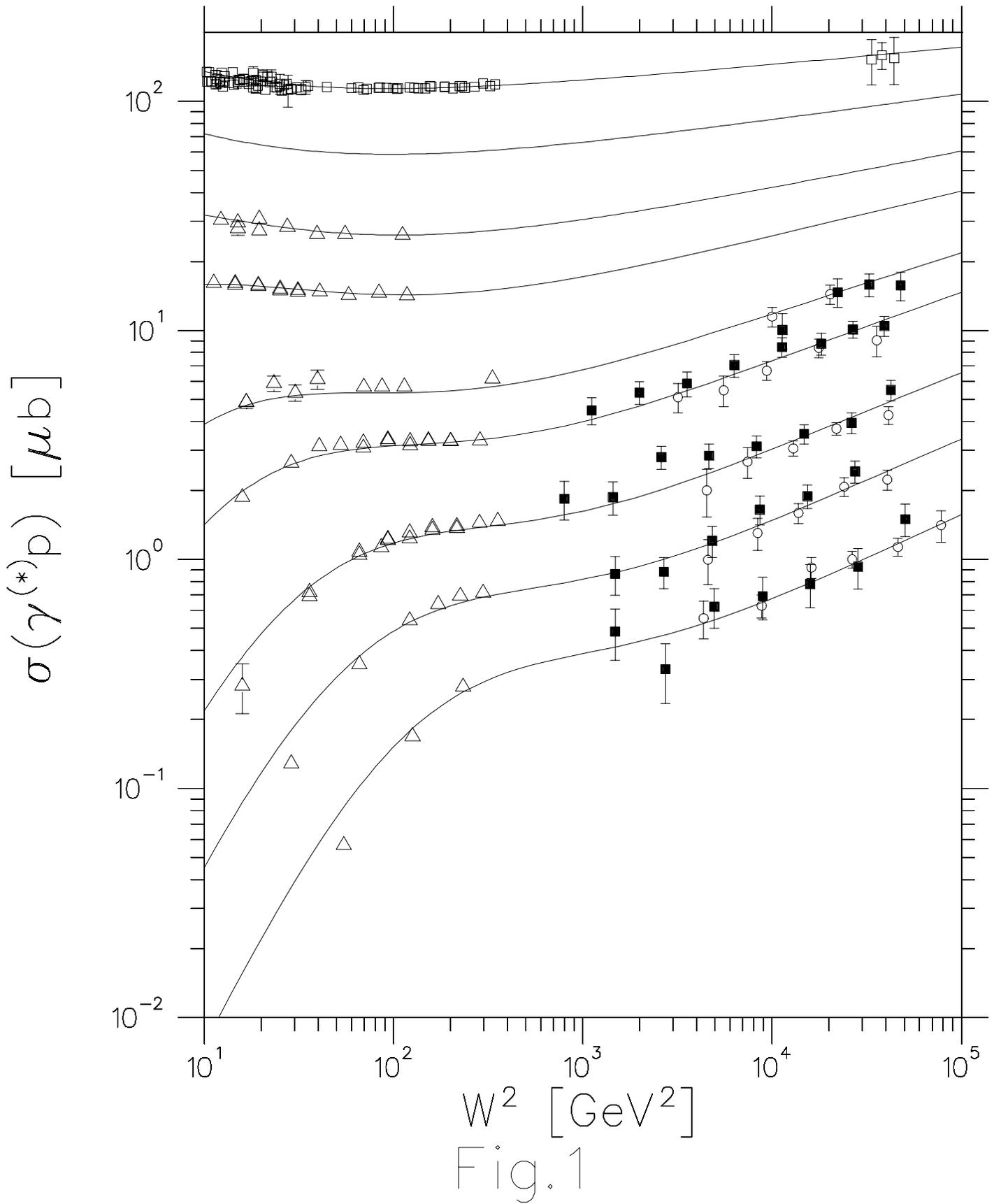
\end{document}